\documentclass{elsart}
\newcommand{\be}{\begin{eqnarray}}
\newcommand{\ee}{\end{eqnarray}}

\usepackage{amssymb,psfig}

\begin{document}

\begin{frontmatter}



\title{Critical Tsallis exponent in heavy ion reaction}


\author{Klaus Morawetz}

\address{Max Planck Institute for the Physics of Complex Systems,
N\"othnitzer Str. 38, 01187 Dresden, Germany}

\begin{abstract}
The numerical solution of the nonlocal kinetic equation allows to
simulate heavy ion reactions around Fermi energy. The expansion
velocity and density profile show specific radial dependence which
can be described with a Tsallis exponent of $q=5/3$. This might be
considered as an
indication of a phase transition. 
\end{abstract}

\end{frontmatter}


The question whether multifragmentation in heavy ion collisions around
the Fermi energy is associated with a nuclear matter phase transition
has been investigated for decades. From the Van der
Waals equation of state it is obvious that, in order to describe such possible phase transition, we must have a kinetic
equation capable to describe the second virial coefficient, or,
more specifically, the excluded volume and the correlated pressure in equilibrium. While the
equilibrium virial correction including its quantum equivalent 
has been thoroughly investigated, it is astonishing that its
nonequilibrium extension has not been considered. Therefore, we
have 
developed a nonlocal quantum kinetic theory which leads to
an excluded volume as well as to a correlated pressure
\cite{LSM97}. Solving the resulting
kinetic equation demands no more numerical effort than solving the
standard local Boltzmann equation extended by Pauli-blocking
effects. Some experimental features of correlations are described by
this result
\cite{MT00}.

Searching for signals of a possible phase transition during the
reaction is hindered by the over-shading of various reaction channels
and four orders of magnitude later observation. Therefore dynamical
models provide the correct approach to search for appropriate signals. In
\cite{MTP00} it was found that in central heavy ion reactions around
the Fermi energy a non-Hubblean expansion profile appears.
This was associated with a peculiar density profile and it vanishes towards
standard Hubble expansion for higher energies. This has been
attributed to long range correlations which are
typical for fluctuations near a phase transition. 

Since the
heavy ion reaction is a complicated evolution of a correlated finite size
system we might search for a simpler effective description of main
features of the numerical solution. A promising short cut to describe effectively a
statistics including finite size effects is the nonextensive
statistical mechanics, starting from nonextensive entropies, as suggested
by Druyvenstein \cite{D30}, Renyi \cite{R70}, Sharma \cite{SM75} or  
Tsallis \cite{T88}. For a discussion of the kinetics underlying
generalized statistics see \cite{K01}.
Here it will be
shown that both the velocity as well as the density profile can be
associated with an anomalous diffusion of fractional derivative
Fokker-Planck-like equation \cite{BTG00} with a Tsallis exponent
$q=5/3$. 
Since the
latter value represents the border between Gau\ss{} and L\`evy- like
fluctuations \cite{PT99}, we consider it a hint of phase transition.

Within the  (extended) quasi-particle and quasi-classical
approximations \cite{MLS00}
we keep the gradient terms in the scattering integral of the Kadanoff and Baym
equation and
obtain a quantum non-local kinetic equation \cite{MLS00},
\begin{eqnarray}
\!\!\!\!\!\!\!\!{\partial f_1\over\partial t}&+&{\partial\epsilon_1\over\partial k}
{\partial f_1\over\partial r}-{\partial\epsilon_1\over\partial r}
{\partial f_1\over\partial k}
=
\int\! P^-\!\left [ f_3^-f_4^-\bigl(1\!-\!f_1\!-\!f_2^-\bigr)
\!-\! \bigl(1\!-\!f_3^-\!-\!f_4^-\bigr)f_1f_2^-\right ].
\label{e11ni}
\end{eqnarray}
The arguments of the distribution functions and the corresponding ones of
the quasiparticle energies $\epsilon$ are nonlocal,
$f_1\equiv f(k,r,t)$, $f_2^-\equiv f(p,r\!-\!\Delta_2,t)$, $
f_3^-\equiv f(k\!-\!q\!-\!\Delta_K,r\!-\!\Delta_3,t\!-\!\Delta_t)$, and
$f_4^-\equiv f(p\!+\!q\!-\!\Delta_K,r\!-\!\Delta_4,t\!-\!\Delta_t)$.
The scattering probability is the square of the
T-matrix,
$P^-={dp dq\over(2\pi)^5}\delta\left(\epsilon_1\!+\!
\epsilon_2^-\!-\!\epsilon_3^-\!-\!\epsilon_4^-\!-\!2\Delta_E\right)$ $\times
\left|T\!\left(\!\epsilon_1\!+\!\epsilon_2^-\!-\!\Delta_E,k-{\Delta_K
\over 2},p-{\Delta_K\over 2},q,r-\Delta_r,t-{\Delta_t\over 2}\right)
\right|^2\!
$.
All non-local corrections are given by derivatives of the scattering
phase shift \mbox{$\phi={\rm Im\ ln}T(\Omega,k,p,q,r,t)$}
\begin{eqnarray}
\Delta_t&=&{\partial\phi\over\partial\Omega},\ \ \
\Delta_E=-{1\over 2}{\partial\phi\over\partial t},\ \ \
\Delta_K={1\over 2}{\partial\phi\over\partial r},\ \ \
\nonumber\\
\Delta_2&=&{\partial\phi\over\partial p}-{\partial\phi\over\partial q}-
{\partial\phi\over\partial k},\ \ \ \ \
\Delta_3=-{\partial\phi\over\partial k}
,\ \ \ \ \ 
\Delta_4=-{\partial\phi\over\partial k}-{\partial\phi\over\partial q}.
\label{e6}
\end{eqnarray}
The collision is of finite duration
$\Delta_t$. During this time particles can gain momentum and energy
$\Delta_{K,E}$ due to the medium effect on the collision. Three
displacements $\Delta_{2,3,4}$ correspond to the initial and final
positions of two colliding particles/holes.
The numerical values of the shifts calculated with realistic
potentials are available \cite{MLSK98} and lead to simple off-set in the
the algorithm which simulates the collision \cite{MT00}. 
\begin{figure}[h] 
\parbox[h]{14.5cm}{
\parbox[]{7cm}{
\psfig{file=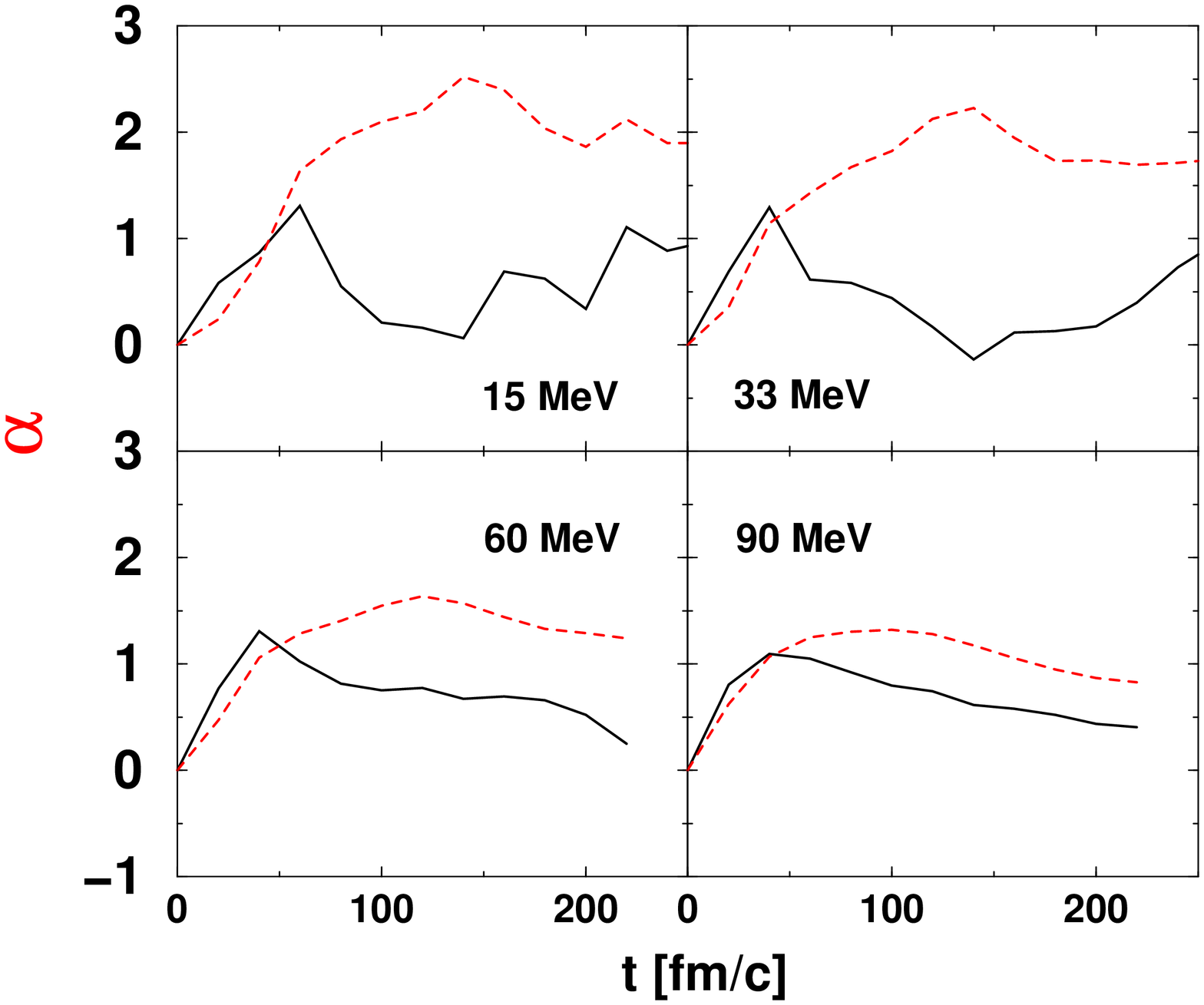,width=7cm,angle=-90}
}
\parbox[]{7cm}{
\psfig{file=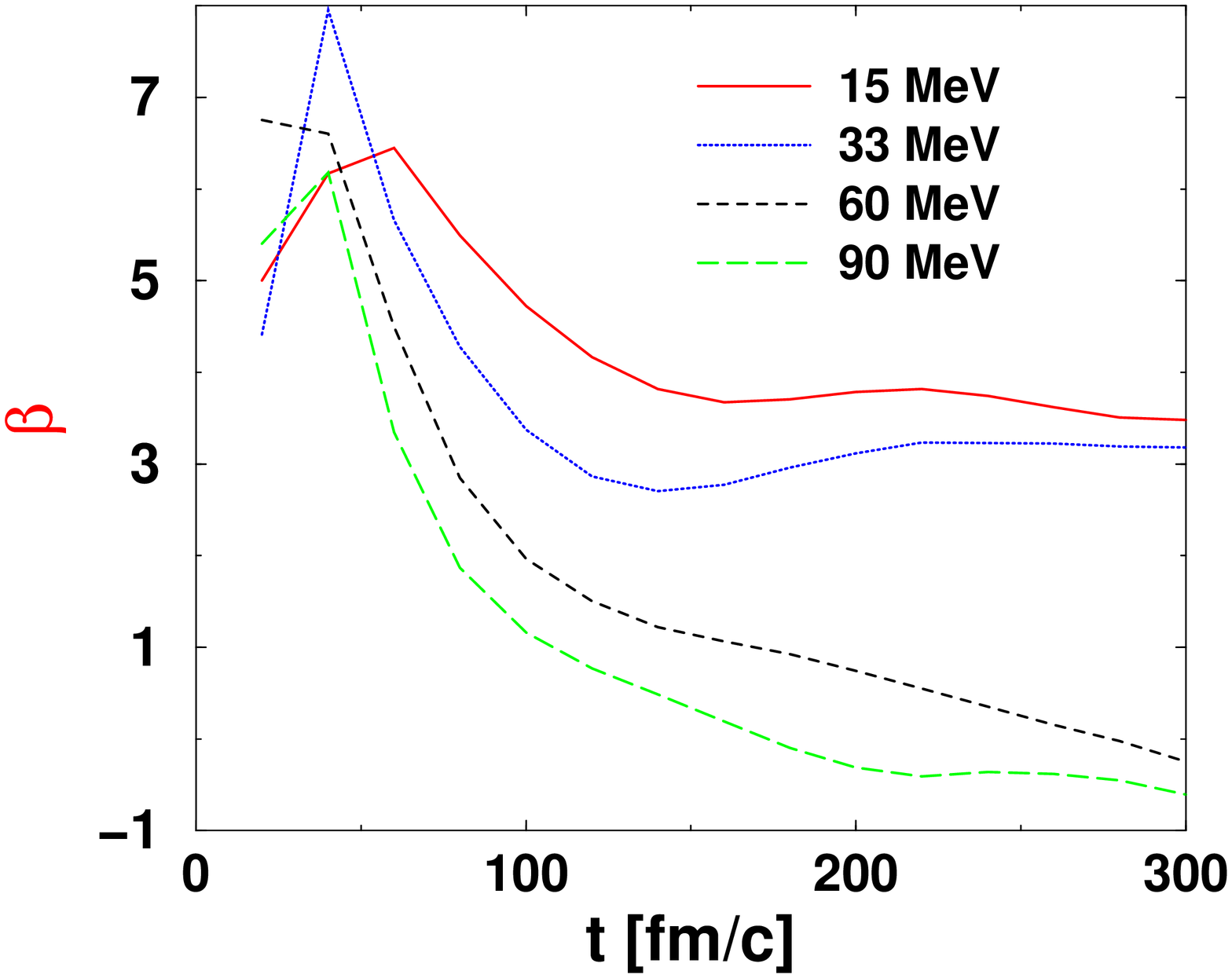,width=7cm,angle=-90}
}}
\caption{The time dependence of the power law velocity 
profile (left) with 
  respect to the radius $v\propto r^\alpha$ for 
different lab
  energies \protect\cite{MTP00}. 
The dotted lines show the surface matter 
  behavior and the solid lines depict the bulk matter 
behavior which was distinguished at the radius $R\sim10$fm.
The time dependence of  
the power law fit of the particle density $n\propto 
r^{-\beta}$ in the surface region
($R > 10$ fm) is seen on the right hand side plot.} 
\label{fig3} 
\end{figure}

We plot  in Fig. 
\ref{fig3} 
the time dependence of the power of radial velocity $v\propto
r^\alpha$ for different bombarding energies and the corresponding
power of radial density dependence in Fig.~\ref{fig3}. 
There are two distinct behaviors dependent on the
bombarding energy. For higher energies, $60$MeV and $90$MeV, we see
that the velocity exponent approaches $\alpha\sim 1$, in agreement with
the Hubble expansion. This is associated with a vanishing density
exponent. The slightly negative values in Fig.~\ref{fig3} can be
understood as ring-like expansions of matter (pancakes). The
interesting observation is that for energies below or around Fermi
energy the density exponent is $\beta\sim 3$ and the surface
velocity exponent $\alpha \sim 2$. This remarkable feature is present in
very different simulations we made with different reactions. It is
more clearly pronounced in the nonlocal scenario when compared to the
local scenario. Therefore, we propose that it is connected with
correlations.

Both exponents can be understood from a
critical Tsallis exponent $q=5/3$. To see this, we first interpret the
radial and time dependent density as a probability distribution in the
sense of a Fokker-Planck equation with fractional derivatives, the solution
of which reads \cite{BTG00}
\be
\!\!\!\!\!\!{\partial \over \partial t} P_\gamma(x,t) =D \nabla^\gamma P_\gamma(x,t)^\nu,
\quad
P_\gamma(x,t)\propto {1\over (t)^{{\gamma+1\over \gamma^2-\gamma+1}}} \left
  [{z^{\gamma(\gamma+1)}\over (1+b z)^{1-\gamma^2}}
\right ]^{1\over 1-2 \gamma},
\label{p}
\ee
with $z={x (|k_1| t) ^{-{\gamma+1\over \gamma^2 -\gamma +1}}}$. The
order of fractional derivative $\gamma$ is linked to the Tsallis
exponent by $q={\gamma+3\over \gamma+1}\approx {5\over 3}$. For a
critical value of $\gamma=2$ or $q=5/3$, one obtains indeed the above
observed cubic density profile
$P_2(x,t,\gamma)\propto x^{-3}$.
The associated velocity profile can be explained also by observing
that for $\gamma=2$ from (\ref{p}) follows
\be
{\partial \over \partial t} P_2(x,t)\propto {1\over t^2} {1\over z^2 (1+b
  z)^2}\qquad z={x (|k_1| t)^{-1}},
\ee
and the velocity profile can be estimated to
\be
v(r) \approx <x {\dot P_2}(x,t)>\propto {1\over b^2} 
(\ln(1+b z)+{1\over 1+b z})\approx{1\over b^2}+{z^{2}\over 2}+o(b z^3).
\ee
This explains the radial quadratic exponent of the observed velocity
profile.

In conclusion, we saw that for bombarding energies below the Fermi energy
there appears a peculiar quadratic velocity profile with respect to
the radius and an associated  
cubic power law of the density. Both features can be
fitted by a fractional derivative Fokker-Planck equation with the
critical Tsallis exponent $q=5/3$. Since this exponent marks exactly
the transition between Gau\ss{}- and L\`evy-like diffusion we suggest
that the observed radial dependence might be associated with a phase
transition during the reaction stage. 





\end{document}